\documentstyle[12pt,fleqn]{article}
\catcode`\@=11
\@addtoreset{equation}{section}

\def\be{\begin{equation}}
\def\ee{\end{equation}}
\def\ba{\begin{eqnarray}}
\def\ea{\end{eqnarray}}

\def\la{\mathrel{\mathpalette\fun <}}
\def\ga{\mathrel{\mathpalette\fun >}}
\def\fun#1#2{\lower3.6pt\vbox{\baselineskip0pt\lineskip.9pt
        \ialign{$\mathsurround=0pt#1\hfill##\hfil$\crcr#2\crcr\sim\crcr}}}

\begin{document}
\begin{titlepage}
\null\vspace{-62pt}
\begin{flushright}FERMILAB-Pub-94/086-A\\
May, 1994 (revised)
\end{flushright}

\vspace{.2in}
\baselineskip 24pt
\centerline{\large \bf{Polynomial Hybrid Inflation}}
%\centerline{\large \bf{ }}

\vspace{.5in}
\centerline{Yun Wang}
\vspace{.2in}
\centerline{{\it NASA/Fermilab Astrophysics Center}}
\centerline{\it Fermi National Accelerator Laboratory, Batavia, IL 60510-0500}

\vspace{.7in}
\centerline{\bf Abstract}
\begin{quotation}

We study a simple extension of Linde's hybrid inflation model, with
the inflaton mass term replaced by the most general renormalizable potential
for $\phi$. The unprocessed power spectrum of density perturbations can have
two minima and one maximum, roughly corresponding to two steep regions

separated by a somewhat flat region in $V(\phi)$.
In the examples studied here, sufficient amount of inflation and normalization

to COBE require a vacuum scale of $10^{16}$ GeV and a $\phi$ mass of $10^{13}$

GeV. Depending on the initial value of $\phi$, our model can give either less

($\sim n<1$) or more ($\sim n>1$) power on small scales compared to the

scale-invariant spectrum ($n=1$), given the normalization to COBE.

\end{quotation}

PACS index numbers: 98.80.Cq

\end{titlepage}

\baselineskip=24pt

\section{I. Introduction}

Inflation has remained the most attractive solution to the problems of the
standard Cosmology, which are the smoothness problem, the flatness problem,

and the formation of structure problem.

Up to date, there has been no definitive model of inflation which solves
all the old problems without creating new ones. The search for more
appealing models of inflation continues.

Recently, Linde proposed ``hybrid inflation''
[\ref{Linde}]. The effective potential of this model is given by
\be
V(\sigma, \phi)= \frac{1}{4\lambda}(M^2-\lambda \sigma^2)^2+\frac{m^2}{2}\,
\phi^2 + \frac{g^2}{2} \, \phi^2 \sigma^2.
\label{eq:Linde}
\ee
This model looks like a hybrid of chaotic inflation with an inflaton

($\phi$) mass term and the usual theory with spontaneous symmetry breaking
with a Higgs field $\sigma$.

Inflation usually ends by the final stage rapid rolling of the
slow-rolling inflaton field $\phi$, or by a first order phase transition.
The important difference between the model of Eq.(\ref{eq:Linde})
and other theories of this type [\ref{KofL}] [\ref{KofP}] [\ref{Katie}] is:
in hybrid inflation, inflation ends with the very rapid rolling of the
scalar field $\sigma$, which is triggered by the inflaton field $\phi$.
The Higgs field $\sigma$ remains a physical degree of freedom after the Higgs
effect in an underlying gauge theory with spontaneous symmetry breaking; hence
it acquires positive values only, which removes the possibility of domain
wall formation in this theory.

Hybrid inflation is very elegant and simple. It gives the appropriate
amplitude of density perturbations with reasonable parameter choices.
However, the raw power spectrum has a minimum on a usually large scale, which
means that there is likely less power on the large scales relevant to
observation. There has always been a lot of interest in obtaining
power spectra with more power on large scales [\ref{Sal}].
We are interested in finding a model of hybrid inflation which

gives a more complex behavior of the power spectrum, with sufficient power

on large scales.

In the spirit of simplicity, we consider the following effective potential:
\be
V(\sigma, \phi)= \frac{1}{4\lambda}(M^2-\lambda \sigma^2)^2+
f_0(\phi) + \frac{g^2}{2} \, \phi^2 \sigma^2,
\label{eq:V}
\ee
where
\be
f_0(\phi)=A \phi^4/4 +B \phi^3/3+ C\phi^2/2.
\label{eq:f0}
\ee
$f_0(\phi)$ is the most general renormalizable potential for $\phi$ with
$f_0(0)=0$. Note that $C\equiv m^2>0$ for a positive $\phi$ mass term.
Hodges et al. studied this potential in the context
of the usual slow-roll inflation [\ref{Hodges}], where they looked for sets
of parameters that can strongly break the scale invariance and give a valley
in the usual Zel'dovich spectrum. In this work we study the potential in
more details within the context of hybrid inflation, and normalize

the power spectrum using the COBE DMR results [\ref{COBEn}].

\section{II. General properties of the inflaton potential}

Consider the effective potential $V(\sigma,\phi)$ of Eq.(\ref{eq:V}). The
effective mass of the $\sigma$ field is
\be
m_{\sigma}^2=\frac{\partial^2 V(\sigma,\phi)}{\partial\sigma^2}
=3\lambda \sigma^2+(g^2\phi^2-M^2).
\ee
For $\phi>\phi_c \equiv M/g$, $\sigma=0$ is the only minimum of

the effective potential in the $\sigma$ direction.

In hybrid inflation, we are interested in potentials which have a much
greater curvature in the $\sigma$ direction than in the $\phi$ direction

[\ref{Linde}].
For sufficiently large $\phi$, this implies $A \ll g^2/3$. When this
condition is satisfied, the $\sigma$ field rolls down to $\sigma=0$ initially,
while the $\phi$ field can remain large for a much longer time.

Define a dimensionless variable
\be
y \equiv \frac{\phi}{\phi_0},
\ee
where $\phi_0$ is a constant with the dimension of mass.
The potential of the $\phi$ field can be written as:
\ba
&&V(0,\phi)= V_0 [1+b f(y)], \nonumber\\
&&f(y)=\frac{y^2}{2}+\frac{\alpha}{3} \, y^3+\frac{\beta}{4}\, y^4,
\label{eq:Vphi}
\ea
with
\ba
&&V_0 \equiv\frac{M^4}{4\lambda}, \hskip 2cm
b \equiv \frac{C\phi_0^2}{V_0}=\frac{m^2\phi_0^2}{V_0},\nonumber\\
&& \alpha\equiv \frac{B \phi_0}{C}, \hskip 2cm
\beta \equiv \frac{A \phi_0^2}{C}.
\nonumber
\ea
Note that $f'(y)=y[1+\alpha y+\beta y^2]$, $f''(y)=1+2\alpha y+3\beta y^2$.
The roots of $f'(y)=0$ are

\ba
&&y=0, \hskip 2cm y_{\pm}=\frac{1}{2\beta}\left[-\alpha\pm
\sqrt{\alpha^2-4\beta} \right], \\
&& \hskip 3cm \left(f''(y_{\pm})=\frac{\sqrt{\alpha^2-4\beta}}{2\beta}
\left[ \sqrt{\alpha^2-4\beta}\mp \alpha\right] \right).\nonumber
\ea
$y=0$ is always a minimum of $f(y)$. If $\alpha^2<4\beta$, $y=0$ is the only

extremum of $f(y)$. If $\alpha^2=4\beta$, $y_+=y_-=-2/\alpha$ is the inflection

point of $f(y)$. If $\alpha^2 > 4\beta$, there are three cases:
(i) $\beta<0$, $y_+<0$, $y_->0$, $y_{\pm}$ are maxima of $f(y)$, $y$ (
i.e. $\phi$) has to tunnel to reach its vacuum if $y>y_-$ initially;

(ii) $\beta>0$ and $\alpha>0$, $y_-<y_+<0$, $y_-$ is a second vacuum
and $y_+$ is the maximum of $f(y)$, there are no extremum for $y>0$;
(iii) $\beta>0$ and $\alpha<0$, $y_+>y_->0$, $y_+$ is the false vacuum
if $\alpha^2<9\beta/2$, $y=y_+$ and $y=0$ are degenerate vacua
if $\alpha^2=9\beta/2$, otherwise $y_+$ is the true vacuum and a positive

constant term must be added to $f(y)$ to yield zero cosmological constant.

It is desirable that we demand $\phi=0$ is the only extremum (the global
minimum) of the $\phi$ potential $V(0$,$\phi)$, which
is the property of the original hybrid inflation model. This allows
us to avoid complications (such as a false vacuum) not of
interest to us here. Hence we want either $\alpha^2<4\beta$, or $\beta>0$ and
$\alpha>0$.

The extrema of $f'(y)$ roughly correspond to the extrema of the power
spectrum, although the power spectrum has an additional minimum

(the only extremum in Linde's model [\ref{Linde}] for which
$f''(y)=1$). Different parameter choices give power spectra with
different features. The roots of $f''(y)=0$ are
\be
y=y_{a,b}\equiv \frac{1}{3\beta} \left[ -\alpha\mp
\sqrt{\alpha^2-3\beta} \right].
\ee
Since $f'''(y_{a,b})=\mp 2\sqrt{\alpha^2-3\beta}$,

for $\alpha^2 >3\beta$, $y_a$ and $y_b$ are
the maximum and minimum of $f'(y)$ respectively.

$\alpha^2=3\beta$ gives the inflection point of $f'(y)$.
For $\alpha^2 >3\beta$, there are three cases:
(i) $\beta<0$, $y_a>0$, $y_b<0$,  but the conditions
for $\phi=0$ being the only extremum of $V(0,\phi)$ are not satisfied;
(ii) $\beta>0$ and $\alpha>0$,

$y_a<y_b<0$; (iii) $\beta>0$ and $\alpha<0$, $y_b > y_a>0$.

If $\alpha^2 <3\beta$ or $\alpha=\beta=0$ (which corresponds to Linde's
model of hybrid inflation), $f''(y)=0$ has no roots; this and case
(ii) of $\alpha^2>3\beta$ closely resemble Linde's model, which we will

not discuss here.

We are interested in case (iii) of $\alpha^2 >3\beta$, for which
\be
 \beta>0, \hskip 1cm \alpha<0; \hskip 2cm
3\beta < \alpha^2<4\beta,
\ee
or
\be
A>0, \hskip 1cm B<0; \hskip 2cm 3AC<B^2 <4AC,
\ee
in terms of the constants $A$, $B$, and $C$ from Eq.(\ref{eq:f0}).

\section{III. The equations of motion}

Setting $\sigma=0$, the equations of motion are
\ba
&&H^2=\frac{8\pi G}{3}\, \left[V(\phi)+ \frac{\dot{\phi}^2}{2}\right],
\nonumber\\
&& \ddot{\phi}+3H\dot{\phi}=-V'(\phi).
\ea
where $V(\phi)\equiv V(0,\phi)$ (see Eq.(\ref{eq:Vphi})).
Inflation occurs if $H^2> 4\pi G \dot{\phi}^2$.

Let us define dimensionless variables
\be
\overline{H} \equiv \frac{H}{H_0}, \hskip 2cm
\tau=H_0 t,
\ee
where $H_0^2 \equiv 8\pi GV_0/3$. The equations of motion can
be written as
\ba
&&\overline{H}^2= 1+b\, f(y)+\frac{1}{2}\, \left(\frac{dy}{d\tau}\right)^2,
\nonumber\\
&& \frac{d^2 y}{d\tau^2}+3\, \overline{H}\frac{dy}{d\tau}=-b \, f'(y),
\ea
where $y\equiv \phi/\phi_0$ with $\phi_0=\sqrt{3/(8\pi)}\,M_{\rm P}$.

The slow-roll conditions [\ref{KT}] are
\be
c_1 \equiv \frac{1}{\sqrt{18}} \left[ \frac{b\,f'(y)}{1+b\, f(y)}\right] \ll 1,
\hskip 2cm
c_2\equiv \frac{1}{9}\left[ \frac{b\, f''(y)}{1+b\, f(y)}\right], \hskip 0.5cm
|c_2| \ll 1.
\label{eq:slow-roll}
\ee
When both conditions are satisfied, we have
\ba
&& \overline{H}^2 \simeq 1+b \,f(y), \nonumber \\
&& 3\, \overline{H}\frac{dy}{d\tau} \simeq -b \, f'(y).
\ea
The number of $e$-folds from the beginning of inflation is
\ba
N(y) &\equiv& \int^{\tau}_0 \overline{H} \,d\tau
\simeq \frac{3}{b} \int^{y_i}_y dy \, \frac{1+b\, f(y)}{f'(y)} \nonumber\\
&=& 3 \left[\frac{1}{b}  \int^{y_i}_y\, \frac{dy}{f'(y)}+
\int^{y_i}_y\, dy\, \frac{f(y)}{f'(y)} \right].
\ea
After simple integration we find
\be
N(y) = 3 \left[ \frac{1}{b} \ln\left(\frac{y_i}{y}\right)+
\frac{1}{b} \mu(y;y_i,\alpha,\beta)+\nu(y;y_i,\alpha,\beta)\right],
\label{eq:N(y)}
\ee
where
\ba
&& \mu(y;y_i,\alpha,\beta)\equiv\frac{1}{2} \ln
\left( \frac{1+\alpha y+\beta y^2}
{1+\alpha y_i+\beta y_i^2}\right)- \frac{\alpha}{
\sqrt{4\beta-\alpha^2} }\, \left[\arctan \left(\frac{\alpha+2\beta y_i}
{\sqrt{4\beta-\alpha^2} }\right) \right. \nonumber\\
&& \left.\hskip 3cm -\arctan \left(\frac{\alpha+2\beta
y}{\sqrt{4\beta-\alpha^2} }\right)
\right], \nonumber\\
&& \nu(y;y_i,\alpha,\beta)\equiv \frac{(y_i-y)}{24 \beta} \left[
3\beta(y_i+y)+2\alpha\right]+\frac{(5\beta-\alpha^2)}{12\beta^2}\,
\mu(y;y_i,\alpha) \nonumber\\
&& \hskip 3cm   +\frac{(\alpha^2-4\beta)}{12\beta^2}\,
\ln \left( \frac{1+\alpha y+\beta y^2}{1+\alpha y_i+\beta y_i^2}\right).
\ea

At $\phi \leq \phi_c\equiv M/g$, the phase transition with symmetry breaking
occurs. The minimum of the $\sigma$ field changes to
\be
\sigma_0(\phi)=\sqrt{\frac{M^2(\phi)}{\lambda}}, \hskip 2cm
{\rm with} \hskip 1cm M^2(\phi)\equiv g^2(\phi_c^2-\phi^2).
\ee
The effective mass of $\sigma$ at $\sigma=\sigma_0$ is
$m^2_{\sigma}(\sigma_0)=\partial^2 V(\sigma_0,\phi)/\partial\sigma^2
=2 M^2(\phi)$. When $\phi$ drops below $\phi_c$,
$\sigma$ rolls down to $\sigma_0$ within $\Delta t
\sim H^{-1}$ if  $m^2_{\sigma}(\sigma_0) \gg H^2$. If the slow-roll conditions
are satisfied, this gives
\be
c_f \equiv  \left(\frac{3}{8}\right)
\frac{y_c^3\,[1+b\, f(y_c)]^2}{b f'(y_c)} \ll \frac{\lambda}{g^2} ,
\label{eq:sigma-roll}
\ee
where $y_c \equiv \phi_c/\phi_0=\sqrt{8\pi/(3g^2)}\,(M/M_{\rm P})$.
$\sigma$ oscillates around $\sigma_0(\phi)$ and loses energy due to
the expansion of the Universe. $\sigma$ cannot simply relax, because
$V(\sigma,\phi)$ has a non-zero slope in the $\phi$ direction at $\sigma
=\sigma_0(\phi)$. $\phi$ rolls down to $\phi=0$ within $\Delta t \sim H^{-1}$
if $\partial^2V(\sigma,\phi)/\partial \phi^2 \gg H^2$ near
$\phi=\phi_c$, which can be written as
\be
c_f \ll 1.
\label{eq:phi-roll}
\ee
The reheating in our model can proceed via the standard way of coherent

oscillations [\ref{KT}].

The amplitude of density perturbations produced during inflation is

[\ref{Lyth}]
\be
\left.\frac{\delta\rho}{\rho}\right|_{RH=k}^{\rm HOR}=
\left.\frac{C_{\rho}\,H^2}{2\pi \dot{\phi}}\right|_{RH=k}
\left[ 1+(2\gamma-1)\epsilon+\gamma \delta\right],
\label{eq:drho}
\ee
where $C_{\rho}=-6/5, -4/3$ depending on when the density perturbations
re-enter the horizon (during matter-dominated or radiation-dominated era).
$R$ is the scale factor, $k$ is the wavenumber,
$\gamma=2-\ln 2-\gamma' \simeq 0.7296$ with $\gamma'$ denoting Euler's
constant, and
\be
\epsilon\equiv -\frac{\dot{H}}{H^2}\simeq 3 c_1^2, \hskip 2cm
\delta\equiv \frac{\ddot{\phi}}{H\dot{\phi}}\simeq -3 c_2,
\ee
where we have used the slow-roll conditions in relating $\epsilon$
and $\delta$ to $c_1$ and $c_2$.
Eq.(\ref{eq:drho}) is a second
order expression for the unprocessed power spectrum, with
$\epsilon$ and $\delta$ characterizing corrections to the usual first order

expression ($\epsilon=\delta=0$). Ref.[\ref{Rocky}] shows that these

corrections may be important in relating spectral indices to tensor and scalar

amplitudes in inflation.

In terms of our dimensionless parameters,
\be
\label{eq:drho-P0}
\left.\frac{\delta\rho}{\rho}\right|_{RH=k}^{\rm HOR}= P_0\,
\left(\frac{\delta\rho}{\rho}\right)_*\equiv
P_0\, \left(\frac{\delta\rho}{\rho}\right)_0
\left[ 1+(2\gamma-1)\epsilon+\gamma \delta\right],
\ee
where

\be
P_0 \equiv \left(\frac{3|C_{\rho}|g^2}{4\pi \sqrt{\lambda}}\right)y_c^2,
\ee
roughly sets the amplitude of density fluctuations, while
\be
\left(\frac{\delta\rho}{\rho}\right)_0 \equiv \frac{\overline{H}^2}{3
\left|\frac{{\rm d}y}{{\rm d}\tau}\right|}\simeq
\frac{[1+b\, f(y)]^{3/2}}{b\,f'(y)}\equiv F(y),
\ee
determines the shape of the power spectrum.

There are four independent parameters in our model, ($M$, $m$, $\alpha$,
$\beta$) or ($y_c$, $b$, $\alpha$, $\beta$). Fitting two
physical observations, the minimum amount of inflation

and the anisotropy on large scales from COBE DMR observations,
we are left with two free parameters (see Sec.V). It turns out
most convenient for us to choose $y_c$ and $b$
to be the free parameters.

\section{IV. Power spectrum shape function $F(y)$}

The function $F(y)$ determines the shape of the raw power spectrum.
To study the roots of $F'(y)=0$ analytically, let us define
\be
x\equiv \sqrt{K}\, y, \hskip 1cm
b_0 \equiv \frac{b}{K}, \hskip 1cm
\alpha_0\equiv \frac{\alpha}{\sqrt{K}}, \hskip 1cm
\beta_0 \equiv \frac{\beta}{K}.
\label{eq:x-y}
\ee
We can write
\ba
&&f(y)=\frac{w(x)}{K}, \hskip 1cm w(x) \equiv\frac{x^2}{2}+\frac{\alpha_0}{3}
 \, x^3+\frac{\beta_0}{4}\, x^4, \nonumber\\
&& F(y)=\left(\frac{\sqrt{K}}{b}\right)W(x), \hskip 1cm
W(x) \equiv \frac{[1+b_0\, w(x)]^{3/2}}{w'(x)}.
\ea
Since $F'(y)=W'(x)/b_0$, the roots of $F'(y)=0$ are related
to the roots of $W'(x)=0$ as follows
\be
y_{1,2,3}=x_{1,2,3}/\sqrt{K}.
\ee
We can choose $K$ such that one of the roots ($x_{a,b}$) of $w''(x)=0$ is 1.
This gives us a relation between $\alpha_0$ and $\beta_0$:
\be
3\beta_0+2\alpha_0+1=0.
\ee
Let us choose $x_b=1$, i.e., we scale the $\phi$ field with its value at the
minimum of
$w'(x)$. This gives us (see Sec.II)
\[
x_b= \frac{1}{3\beta_0} \left[ |\alpha_0|+ \sqrt{\alpha_0^2-3\beta_0}
\right]=1,
\]
i.e.,
\be
K=\left[ |\alpha|-\sqrt{\alpha^2-3\beta}\right]^2=
\left(\frac{|\alpha|}{2-\Delta}\right)^2,
\label{eq:K}
\ee
where we have defined $\Delta \equiv \alpha_0+2$. Hence
\be
\beta=\frac{\alpha^2}{3}\, \left[1-\left(\frac{1-\Delta}{2-\Delta}\right)^2
\right].
\ee
Now we have
\be
0< \Delta < 1,  \hskip 1cm \frac{1}{3}<\beta_0<1,
\hskip 2cm {\rm for} \,\,\,
x_a=\frac{1}{3-2\Delta}, \hskip 1cm x_b=1.
\ee
where the upper bound on $\Delta$ comes about because $x_a <x_b$, and we
have chosen $x_b=1$.

$\Delta=0$ corresponds to the inflection point of $w(x)$, and
$\Delta=1$ the inflection point of $w'(x)$.

$W(x)$ is determined by $\Delta$ and $b_0$.

Fig.1 shows $w(x)=K \, f(y)$ for $\Delta=0.1$ and $0.5$.
Fig.2 shows the corresponding $W(x)=(b/\sqrt{K}) F(y)$ for
$b_0=0.1$, $1$, and $10$.

We have chosen $V(\phi)=V_0[1+b_0w(x)]$ to increase monotonically
with $\phi$, i.e., $w(x)$ increases with $x$. Since $1/w'(x)$ has a minimum at
$x_a=1/(3-2\Delta)$ and a maximum at $x_b=1$, we expect $W(x)$ to
have a minimum at $0< x_1 \la x_a$, a maximum at $x_2 \ga x_b $, and
a second minimum at $x_3> x_2$.
Note that
\ba
&&W'(x)=\frac{[1+b_0w(x)]^{1/2}}{2[w'(x)]^2} \,
\left\{ 3b_0[w'(x)]^2-2w''(x)[1+b_0w(x)] \right\}; \\
&&W''(x)=\frac{[1+b_0w(x)]^{1/2}}{[w'(x)]^2} \,
\left\{ 2b_0w'(x) w''(x)-2w^{(3)}(x)[1+b_0w(x)] \right\}, \nonumber\\
&& \hskip 4cm {\rm at}\,\, W'(x)=0.\nonumber
\ea
$W'(x)=0$ can be solved numerically for given $\Delta$ and $b_0$.

Fig.3 shows the three roots $x_{1}<x_2<x_3$ as functions of $b_0$ for
$\Delta=0.1$, $0.5$, and $0.9$.

To understand the behavior of $x_{1,2,3}(b_0)$ qualitatively, let us
solve $W'(x)=0$ in various limits.
$W'(x)=0$ can be written as $3b_0[w'(x)]^2-2w''(x)[1+b_0w(x)]=0$, or
\be
12(1+2\alpha_0 x+3\beta_0 x^2)=b_0x^2[9\beta_0^2 x^4+18\alpha_0\beta_0 x^3+5(
2\alpha_0^2+3\beta_0)x^2+20 \alpha_0 x+12] \, .
\label{eq:W'(x)=0}
\ee
For $b_0\gg 1$, $x_1 \ll 1$. We can drop all except the last term

on the right hand side of Eq.(\ref{eq:W'(x)=0}), which gives
\be
x_1 \simeq \frac{\sqrt{b_0+(1-\Delta)^2}+\Delta-2}{b_0-3+2\Delta},
\hskip 2cm b_0 \gg 1.
\ee
For $b_0\ll 1$, we can drop all the terms on the right hand side of

Eq.(\ref{eq:W'(x)=0}), which gives
\be
x_1 \simeq \frac{1}{3-2\Delta}=x_a, \hskip 1cm
x_2 \simeq 1, \hskip 2cm b_0\ll 1.
\ee
Let $x_2=1+r$ with $r<1$, we can expand all terms in Eq.(\ref{eq:W'(x)=0}) to
second order in $r$. We find
\be
x_2 \simeq 1+\frac{2b_0\left[ \sqrt{1+\omega\Delta^2}-1\right]}
{\omega[12+b_0(1+2\Delta)](1-\Delta)}, \hskip 1cm\omega\equiv
\frac{6b_0[b_0+4(3-2\Delta)]}{[12+b_0(1+2\Delta)]^2(1-\Delta)^2}.
\label{eq:x2}
\ee
Eq.(\ref{eq:x2}) applies quite well for all $b_0$, with $\Delta$ not
far from the inflection point $\Delta=0$ (say, $0<\Delta \la 0.5$).

$W'(x)=0$ can be written as
\ba
&&(x-1)\left\{ 12[x(3-2\Delta)-1]-b_0x^2[ \omega(x)-2x(6x^2-15x+10)\Delta
-4x^2(x-2)\Delta^2]\right\} \nonumber\\
=&& 2b_0x^4(2x-3)^2 \Delta^2,
\ea
where
$\omega(x) \equiv 9x^3-27x^2+28x-12$. For $b_0\rightarrow \infty$ and

$\Delta=0$, the above equation becomes
\be
(x-1)\, \omega(x)=0,
\ee
which has the exact solution $x=1$ and $x=x_*\equiv 1.5447884$ [\ref{Hodges}].

To order $\Delta$, the second minimum of $W(x)$ is

\be
x_3 \simeq x_*\left[ 1+\frac{2(6x_*^2-15x_*+10) \Delta}
{27x_*^2-54x_*+28} \right] \simeq  x_*\left(1+
\frac{\Delta}{4}\right),
\hskip 1.5cm b_0\rightarrow \infty.
\ee
For $b_0$ large and to order $\Delta$,

\be
x_3\simeq x_*\left[1+\frac{\Delta}{4}+ \frac{5.25-2.76 \Delta}
{b_0(4+2.315\Delta)-6.7+2.76\Delta }\right], \hskip 2cm b_0\gg 1.
\ee

For $b_0\ll 1 $, $x_3 \gg 1$. Keeping the highest power in $x$ terms from
each side in Eq.(\ref{eq:W'(x)=0}), we find
\be
x_3 \simeq \left[\frac{12}{b_0(3-2\Delta)}\right]^{1/4}, \hskip 2cm b_0\ll 1.
\ee

\section{V. Sufficient inflation and normalization to COBE}

The minimum amount of inflation required to solve the smoothness problem is
[\ref{KT}]
\be
N_{\rm min}=60.6+\frac{2}{3}\, \ln y_c+\ln \overline{H}(y_i)+
\frac{1}{3}\, \ln\left( \frac{T_{\rm RH}}{10^{10}\,{\rm GeV}}\right).
\ee
$y_i$ and $y_c$ are the scaled $\phi$ field at the beginning and end of
inflation respectively. $T_{\rm RH}$ is the reheating temperature.

During inflation, the comoving scale $\lambda$ which crosses outside the

horizon when $\phi=y \phi_0$ is given by
\be
\ln \left(\frac{\lambda}{h^{-1}{\rm Mpc}}\right) \simeq 8-N(y)+(
N_{\rm tot}-N_{\rm min}),
\ee
where $N(y)$ is the number of $e$-folds of expansion since the beginning of

inflation, and $N_{\rm tot}$ denotes the total amount of inflation.

We can choose $y_i >y_3$, or $y_2>y_i>y_1$, such that a valley appears in
the power spectrum at an interesting scale. At the end of inflation
\be
N(y_c;b,\alpha,\beta)=N_{\rm tot}.
\ee
For given $(\alpha,\beta)$ and $N_{\rm tot}$, this gives $y_c$ in terms of $b$.

The processed power spectrum is

\be
P(k,t_0)=\frac{\pi}{(2\pi)^3} \left[\frac{d_{\rm H}(t_0)}{R(t_0)}\right]^4
\, \left[ \left(\frac{\delta\rho}{\rho}\right)^k_{\rm HOR}\right]^2\,
T^2(k) \, k.
\ee
$T(k)$ is the transfer function. For cold dark matter, we use [\ref{Bardeen}]
\be
T(q)=\frac{\ln(1+2.34q)}{2.34q} \left[1+3.89q+ (16.1q)^2+
(5.46q)^3+(6.72q)^4\right]^{-1/4},
\ee
where $q \equiv k/(\Omega_0 h^2 {\rm Mpc}^{-1})$. We take $\Omega_0=1$,

$h=0.5$, the standard CDM model.

Cosmic anisotropy produces an excess variance $\sigma^2_{\rm sky}$
in the $\Delta T$ maps produced by the Differential Microwave Radiometer

(DMR) on COBE that is over and above the instrument noise.

After smoothing to an effective resolution of $10^{\circ}$, this excess,
$\sigma_{\rm
sky}(10^{\circ})$, provides an estimate for the amplitude of the

primordial density perturbation
power spectrum with a cosmic uncertainty of only $12\%$ [\ref{COBEn}].
Following the notation of Ref.[\ref{Josh}], we write

\be
\sigma_T^2(10^{\circ}) =\sum_{l\geq 2} \frac{2l+1}{4\pi}\,
\langle |a_{lm}|^2\rangle\, \exp[-l(l+1)\theta_0^2],
\ee
where $\theta_0=0.425\, \theta_{\rm FWHM}=4.25^{\circ}$ is the Gaussian
angle corresponding to the antenna beam and additional smearing of the raw
data.
$\sigma_T(10^{\circ})$ is directly related to the power spectrum $P(k)$.
For the anisotropy in the cosmic microwave background radiation on large

angular scales, the Sachs-Wolfe effect dominates. For adiabatic perturbations
\be
\langle |a_{lm}|^2\rangle=\frac{H_0^4}{2\pi}\, \int^{\infty}_0
\, {\rm d}k\, k^{-2} P(k)\, j_l^2(kr_{\rm rec}),
\ee
where $j_l$ is a spherical Bessel function and $r_{\rm rec}\simeq
2H_0^{-1}$ is our comoving distance from the surface of recombination.

Based on the latest results of COBE DMR [\ref{COBEn}], we take

$\sigma_T(10^{\circ})=(1.25\pm 0.2) \times 10^{-5}$ [\ref{Staro}].

For given $(\alpha,\beta)$, this enables
us to normalize the power spectrum $P(k)$, thus determine $b$.

Let us take $\Delta=0.1$ to study the properties of our model.
Fig.4 shows the number of $e$-folds $N(y)$ since the beginning of inflation,

for (i) $\alpha=-2$, $y_i=y_2-0.2$; (ii) $\alpha=-2$, $y_i=y_3+5$; (iii)
$\alpha=-0.8$,
$y_i=y_3+5$.  The $\phi$ field rolls slowly if it starts
in the not too steep region of the potential, i.e., between the first

and second minimum of the power spectrum shape function $F(y)$,

$y_1<y_i\equiv \phi_i/\phi_0<y_3$ (see Fig.2).
For large $\phi_i$ ($y_i>y_3$), $\phi$ rolls more slowly in the

astrophysically relevant region ($N(y) \la 20$) for smaller $|\alpha|$.
Sufficient inflation can be obtained for initial values of

$y\equiv\phi/\phi_0$ greater than the first minimum $y_1$ of the power

spectrum shape function $F(y)$ (see Fig.2).

Fig.5 shows $(\delta\rho/\rho)_{*}=P_0^{-1}(\delta\rho/\rho)_{\rm HOR}$

(see Eq.(\ref{eq:drho-P0}))
corresponding to the parameter choices of Fig.4, to the first

(dash line) and second order (solid line) in slow roll approximation. The
deviation of the first order curve from the second order curve

increases as $\phi$ rolls faster.

Fig.6 gives four examples of the power spectrum $P(k)$ with $\Delta=0.1$.

For $\alpha=-2$, $y_i=y_2-0.2$ (short dash line), $y_2+0.1$ (dot line),
and $y_3+5$ (long dash line) respectively. For $\alpha=-0.8$, $y_i=y_3+5$
(dot - long dash line). The solid line is CDM with $n=1$.
We have used the second order

formula for $\delta\rho/\rho$ in calculating $P(k)$.
If $\phi$ starts by rolling over the peak in the power
spectrum shape function $F(y)$, our model gives somewhat more power
on small scales than the standard CDM with $n=1$.
If $\phi$ starts by rolling down one of the valleys in the power
spectrum shape function $F(y)$, our model gives less power
on small scales than the standard CDM with $n=1$.
$\phi$ rolling down the steeper valley in $F(y)$ at $y>y_3$ leads to more
dramatic reduction of power on small scales.
For given $\Delta$ and $(y_i-y_3)>0$, larger $|\alpha|$ seems to lead to
less power on small scales. But the slow-roll conditions (see
Eq.(\ref{eq:slow-roll})) may break down near $y=y_3$, hence it may not be
appropriate to
use Eq.(\ref{eq:drho}) in the

calculation of $\delta\rho/\rho$. For $|\alpha|\ga 10$ (with $\Delta=0.1$ and

$y_i=y_3+5$), the slow-roll parameter
$c_2>1$ near $y=y_3$. The proper mapping of the entire allowed parameter range
and the
discussion of the tensor contribution to the CMBR anisotropy
will be presented elsewhere [\ref{next}].

The $\sigma$ vacuum scale $M$ and the $\phi$ mass $m$ can be expressed as
\be
\frac{M}{M_{\rm P}}=\sqrt{\frac{3g^2}{8\pi}}\, y_c,
\hskip 1.5cm \frac{m}{M_{\rm P}}=\sqrt{\frac{2\pi b}{3\lambda}}\,
\left(\frac{M}{M_{\rm P}}\right)^2.
\ee
For $\Delta=0.1$ and $\alpha=-2$, $M \sim 10^{16}$ GeV, $m\sim 10^{13}$ GeV.
For given $\alpha$ and $(y_i-y_3)>0$, $P(k)$ is not sensitive to $\Delta$;
the scales $M$ and $m$ vary less than an order of magnitude as we change

$\Delta$ from $0.1$ to $0.9$ (recall that $0<\Delta<1$).
Larger $|\alpha|$ seems to lead to smaller $M$ and $m$ ($M\la 10^{15}$ GeV,

$m\la 10^{12}$ GeV). We have taken the dimensionless coupling constants
$\lambda$ and $g$ to be of order $1$.

Let us now examine the conditions for successful hybrid inflation.
For inflation to occur, we want
$\sigma$ to roll down to $\sigma=0$ initially while $\phi$ remains large,
which requires $A \ll g^2/3$ (see Sec.II). This can be expressed as
\be
\alpha \ll g \left(\frac{M_{\rm P}}{m} \right).
\ee
The above condition is always satisfied since $M_{\rm P}\gg m$.
The fast-roll conditions Eqs.(\ref{eq:sigma-roll}) and (\ref{eq:phi-roll})
are automatically satisfied, because $y_c\ll 1$ in our model.
Typically $c_f \la 10^{-5}$.
This means that both $\sigma$ and $\phi$ roll down to their

true vacuum states almost instantly at the time of the phase transition

($\phi=\phi_c\equiv M/g$), thus putting an abrupt end to inflation (see

Sec.III).

To sum up, the requirement of sufficient inflation and normalization to
COBE give us two constraints on the four
parameters ($b$, $y_c$, $\alpha$, $\beta$) in our model.

($\alpha$, $\beta$) roughly determine the shape of the unprocessed power
spectrum. For given ($\alpha$, $\beta$), we can find ($b$, $y_c$) or
($m$, $M$) which satisfy the physical constraints. The same choice of
parameters also guarantees successful hybrid inflation.
The typical values of the vacuum scale $M$ and the $\phi$ mass $m$
are: $M \la 10^{16}$ GeV, $m \la 10^{13}$ GeV. The dimensionless
constants $\alpha$ and $\beta$ in the scaled $\phi$ potential $f(y)$ (see

Eq.(\ref{eq:Vphi})) are usually of order $1$. The constants in the

original form of the $\phi$ potential (see Eq.(\ref{eq:V})) are:
$A \sim (m/M_{\rm P})^2$, $B \sim m^2/M_{\rm P}$,  $C = m^2$; $\lambda$
and $g$ can be taken to be of order $1$.

\section{VI. Remarks}

The model we have presented here has a very simple potential. It involves
only one extra term, the cubic $\phi$ self-coupling term, compared to

previous theories of its type [\ref{Linde}] [\ref{Sal}].

Inflation can end abruptly in our model.
Sufficient inflation is easily obtained. The power spectrum shape function

$F(y)\propto (\delta\rho/\rho)_{\rm HOR}$ has two valleys and one peak,
while the original hybrid inflation model gives only one valley.

When normalized to COBE, $\phi \equiv y \phi_0$ starting on a
slope in $F(y)$ and going down either of the valleys in $F(y)$

(see Fig.2) leads to less power on small scales ($\sim n<1$), while $\phi$

starting between the peak and the second valley in $F(y)$ (see Fig.2)
leads to more power on small scales

($\sim n>1$), compared to the standard CDM with $n=1$.
Our model obtains interesting power spectra quite naturally.

\vskip 0.2in
\centerline{\bf Acknowledgments}
I thank Rocky Kolb for inspiring conversation and helpful suggestions.
I thank Josh Frieman for helpful comments.
I thank Scott Dodelson and Igor Tkachev for useful discussions.
This work was supported by the DOE and NASA under Grant NAGW-2381.

\newpage
\frenchspacing
\parindent=0pt

\centerline{{\bf References}}

\begin{enumerate}

\item\label{Linde} A. Linde, Phys. Rev. D, {\bf 49}, 748 (1994).

\item\label{KofL} L.A. Kofman and A.D. Linde, Nucl. Phys. {\bf B282}, 555
(1987).

\item\label{KofP}  L.A. Kofman and D.Yu. Pogosyan, Phys. Lett. B {\bf 214}, 508
(1988); D.S. Salopek, J.R. Bond, and J.M. Bardeen, Phys. Rev. D {\bf 40},
1753 (1989); L.A. Kofman, Phys. Scr. {\bf T36}, 108 (1991).

\item\label{Katie} F.C. Adams and K. Freese, Phys. Rev. D., {\bf 43},
353 (1991).

\item\label{Sal} D.S. Salopek, J.R. Bond, and J.M. Bardeen, Phys. Rev. D,
{\bf 40}, 1753 (1989).

\item\label{Hodges} H. Hodges, G. Blumenthal, L. Kofman, and J. Primack,
Nucl. Phy., {\bf B335}, 197 (1990).

\item\label{COBEn} E.L. Wright, G.F. Smoot, A. Kogut, G. Hinshaw, L. Tenorio,
C. Lineweaver, C.L. Bennett, P.M. Lubin, Ap.J. {\bf 420}, 1 (1994);
C.L. Bennett, A. Kogut, G. Hinshaw, {\it et al.}, astro-ph/9401012.

\item\label{KT} E.W. Kolb and M.S. Turner, The Early Universe
(Addison-Wesley Publishing Company, 1990).

\item\label{Lyth} E.D. Stewart and D.H. Lyth, Phys. Lett. B, {\bf 302},
171 (1993).

\item\label{Rocky} E.W. Kolb and S.L. Vadas, FERMILAB-Pub-94/046-A,
astro-ph/9403001 (1994).

\item\label{Bardeen} J.M. Bardeen, J.R. Bond, N. Kaiser, A.S. Szalay, Ap. J.

{\bf 304}, 15 (1986).

\item\label{Josh} F.C. Adams, J.R. Bond, K. Freese, J.A. Frieman, and
A.V. Olinto, Phys. Rev. D, {\bf 47}, 426 (1993).

\item\label{Staro} P. Peter, D. Polarski, and A.A. Starobinsky, DAMTP-R94/20,
(1994).

\item\label{next} Y. Wang, in preparation (1994).

\end{enumerate}

\newpage
\nonfrenchspacing
\parindent=20pt

\centerline{{\bf Figure Captions}}

\vspace{0.2in}

Fig.1. $w(x)=K \, f(y)$ for $\Delta=0.1$ and $0.5$ (dash line).
$f(y)$ is the scaled $\phi$ potential.
\vskip 0.1in

Fig.2. $W(x)=(b/\sqrt{K}) F(y)$ for $\Delta=0.1$ and $0.5$ (dash line), with
$b_0=0.1$, $1$, and $10$. $F(y) \propto (\delta\rho/\rho)_{\rm HOR}$ is the

power spectrum shape function.
\vskip 0.1in

Fig.3. The three roots of $W'(x)=0$, $x_{1,2,3}=\sqrt{K}\, y_{1,2,3}$,
as functions of $b_0$ for $\Delta=0.1$ (solid line), $0.5$ (dash line) and
$0.9$ (dot line).
Note that $x_1<x_2<x_3$ for a given $\Delta$.

\vskip 0.1in

Fig.4. The number of $e$-folds $N(y)$ since the beginning of inflation
with $\Delta=0.1$, for (i) $\alpha=-2$, $y_i=y_2-0.2$; (ii) $\alpha=-2$,
$y_i=y_3+5$; (iii)
$\alpha=-0.8$, $y_i=y_3+5$.

\vskip 0.1in

Fig.5. $(\delta\rho/\rho)_*$ with $\Delta=0.1$, to the first (dash line) and

second order (solid line) in slow roll approximation, for (i) $\alpha=-2$,

$y_i=y_2-0.2$; (ii) $\alpha=-2$, $y_i=y_3+5$; (iii)  $\alpha=-0.8$,
$y_i=y_3+5$.

\vskip 0.1in

Fig.6. Four examples of the power spectrum $P(k)$ with $\Delta=0.1$.

For $\alpha=-2$, $y_i=y_2-0.2$ (short dash line), $y_2+0.1$ (dot line),
and $y_3+5$ (long dash line) respectively. For $\alpha=-0.8$, $y_i=y_3+5$
(dot-long dash line). The solid line is CDM with $n=1$.

\end{document}